\documentclass[amsfonts, amssymb, amsmath, aps, 10pt, pra, twocolumn, superscriptaddress, english, floatfix]{revtex4-2}
\usepackage[utf8]{inputenc}
\usepackage[english]{babel}
\usepackage[T1]{fontenc}

\usepackage{mathdots}
\usepackage{bm}
\usepackage{graphicx}
\usepackage{physics}
\usepackage{dsfont}
\usepackage{mathtools}
\usepackage[dvipsnames]{xcolor}
\usepackage{hyperref}

\hypersetup{
    colorlinks = true,
    linkcolor = RoyalBlue,
    citecolor = RoyalBlue,
    urlcolor = RoyalBlue
}

\graphicspath{{figures/}}

\begin{document}

\title{Long-Distance Genuine Multipartite Entanglement \\ between Magnetic Defects in Spin Chains}

\author{Mirko Consiglio}
\affiliation{Department of Physics, University of Malta, Msida MSD 2080, Malta}

\author{Jovan Odavi\'c}
\affiliation{INFN, Sezione di Napoli, Italy}
\affiliation{Dipartimento di Fisica ‘Ettore Pancini’, Università degli Studi di Napoli Federico II, Via Cintia 80126, Napoli, Italy}

\author{Riccarda Bonsignori}
\affiliation{Institute of Theoretical and Computational Physics, Graz University of Technology,
Petersgasse 16, 8010 Graz, Austria}

\author{Gianpaolo Torre}
\affiliation{Institut Ru\dj er Bo\v{s}kovi\'c, Bijeni\v{c}ka cesta 54, 10000 Zagreb, Croatia}

\author{Marcin Wie\'sniak}
\affiliation{Institute of Theoretical Physics and Astrophysics, Faculty of Mathematics, Physics, and Informatics, University of Gda\'nsk, 80-308 Gda\'nsk, Poland}
\affiliation{International Centre for Theory of Quantum Technologies, University of Gda\'nsk, 80-308 Gda\'nsk, Poland}

\author{Fabio Franchini}
\affiliation{Institut Ru\dj er Bo\v{s}kovi\'c, Bijeni\v{c}ka cesta 54, 10000 Zagreb, Croatia}

\author{Salvatore M. Giampaolo}
\email{sgiampa@irb.hr}
\affiliation{Institut Ru\dj er Bo\v{s}kovi\'c, Bijeni\v{c}ka cesta 54, 10000 Zagreb, Croatia}

\author{Tony J. G. Apollaro}
\affiliation{Department of Physics, University of Malta, Msida MSD 2080, Malta}

\begin{abstract}

We investigate the emergence and properties of long-distance genuine multipartite entanglement, induced via three localized magnetic defects, in a one-dimensional transverse-field XX spin-$1/2$ chain. 
Using both analytical and numerical techniques, we determine the conditions for the existence of bound states localized at the defects. 
We find that the reduced density matrix (RDM) of the defects exhibits long-distance genuine multipartite entanglement (GME) across the whole range of the Hamiltonian parameter space, including regions where the two-qubit concurrence is zero.
We quantify the entanglement by using numerical lower bounds for the GME concurrence, as well as by analytically deriving the GME concurrence in regions where the RDM is of rank two. 
Our work provides insights into generating multipartite entanglement in many-body quantum systems via local control techniques.

\end{abstract}

\maketitle

\section{Introduction}
\label{sect:intro}

Multipartite entanglement is a fundamental resource in a variety of quantum information protocols, including multi-party quantum key distribution~\cite{Epping2017}, quantum teleportation and dense coding~\cite{Yeo2006}, quantum metrology~\cite{Geza2012}, and quantum computation~\cite{Raussendorf2001}. 
Such a centrality in a field so relevant as quantum information has attracted numerous works aimed at its quantification in different conditions. 
While significant progress has been made in characterizing and quantifying multipartite entanglement in pure states --- such as identifying inequivalent classes under {\it Stochastic Local Operations and Classical Communication} (SLOCC) operations~\cite{Du2000, Verstraete2002, Horodecki2024} and developing related measures~\cite{Ma2024} --- studying multipartite entanglement in mixed states remains challenging~\cite{Walter2016, Eltschka2014, Bengtsson2017, Lepori2023, Wiesniak2024} and an open problem. 
A variety of different avenues for tackling mixed-states multipartite entanglement have been proposed, from semidefinite programming~\cite{Jungnitsch2011} to geometric approaches~\cite{Cianciaruso2016} and witnesses~\cite{Sperling2013}. 
By exploiting such quantities, it is now possible to investigate the multipartite entanglement in ground states of many-body systems, and many results for spin-$1/2$ models have already been published over the years~\cite{Guhne2005, Amico2008, Giampaolo2013, Liu2022}. 
Thanks to these advancements, the general behaviors of multipartite entanglement are well known. 
Among them, one of the most relevant is the fact that multipartite entanglement, similar to the bipartite one, is limited in range when evaluated for ground states of local Hamiltonians with short-range interactions. 

Although the spatial range at which two spins have non-zero concurrence can be extended by breaking the translational invariance of the model, e.g. by modifying the edge~\cite{Campos2006, Giampaolo2009, CamposVenuti2007, Giampaolo2010} or the on-site couplings~\cite{Apollaro2006, Plastina2007, Apollaro2008a}, the applicability of a similar approach to extend the range of multipartite entanglement has not been investigated thus far. 
In this paper, we seek to address this gap by addressing the problem of generating long-distance multipartite entanglement among three magnetic defects embedded in the XX spin-$1/2$ Hamiltonian. 
In our approach, the defects are represented by a magnetic field in the transverse direction on three sites different from the rest of the system. 
We investigate the various classes of multipartite entanglement as a function both of the bulk's and the defects' magnetic field intensity and the defects' distance, using a combination of analytical tools, lower bound techniques, and entanglement witnesses. Interestingly, we show how the presence of the magnetic defects defines regions in the Hamiltonian's parameter space where discrete energy eigenstates, associated with localized states, appear in its spectrum. 
These eigenstates can sustain long-distance genuine multipartite entanglement (GME), which can be quantified with GME concurrence \cite{Ma2011}, regardless of the relative distance between the defects. 

The manuscript is organized as follows: in Sec.~\ref{sec:model} we introduce the XX spin-$1/2$ model with magnetic defects and define the regions exhibiting discrete eigenenergies; in Sec.~\ref{sec:Multipartite entanglement} we derive the reduced density matrix (RDM) of the defects' subsystem, and define the figures of merit for tripartite genuine multipartite entanglement that are used in Sec.~\ref{Long-distance GME concurrence} for presenting our main result about long-distance GME concurrence; and in Sec.~\ref{sec:conclusions} we conclude the manuscript and discuss prospects for further research.

\begin{figure}[t]
    \centering 
    \includegraphics[width=\columnwidth]{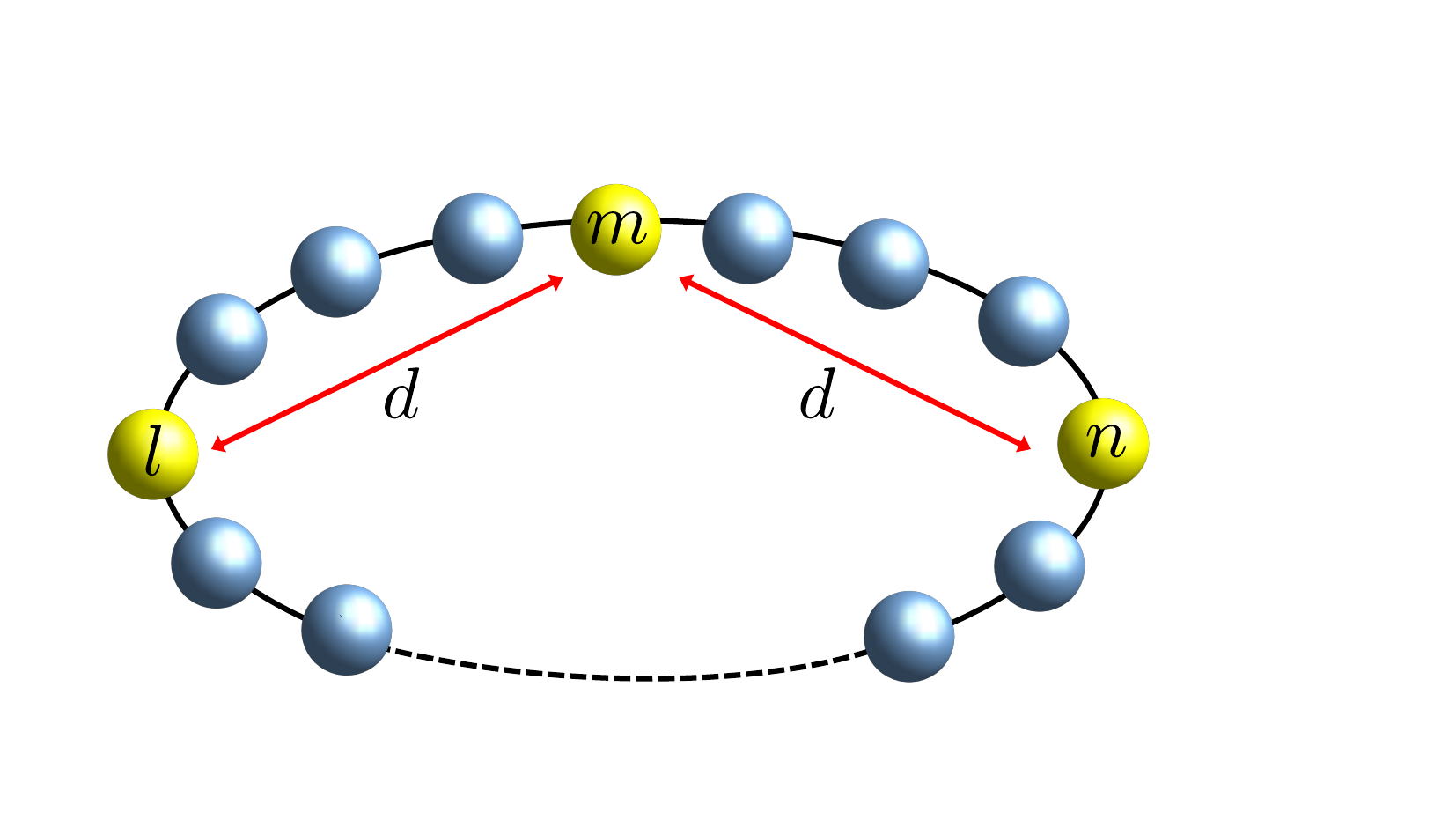}
    \caption{Spin-$1/2$ chain with periodic boundary conditions of even size $N$, described by the Hamiltonian in Eq.~\eqref{eq_model} where on sites $l,m$ and $n$ (the defects) an additional magnetic field $\varepsilon$ is applied. The defects $l$ and $n$ are equidistant from the defect located at site $m$.}
    \label{fig:chain}        
\end{figure}

\section{The model}
\label{sec:model}

We consider the one-dimensional XX spin-$1/2$ chain in a transverse magnetic field $h$, uniform everywhere except for three spins, located at sites $k+jd$ ($j=-1,0,1$), hereafter dubbed defects. 
The Hamiltonian describing the system's dynamics consists of two distinct components:
\begin{equation}
    \label{eq_model}
\hat{H}=\hat{H}_\text{XX}+\hat{H}_\text{def}.
\end{equation}
The first of which is a homogeneous XX Hamiltonian
\begin{equation}
    \label{eq_XX}
    \hat{H}_\text{XX}= \frac{J}{2}\sum_{i=1}^{N}(\hat{\sigma}_i^x\hat{\sigma}_{i+1}^x+\hat{\sigma}_i^y\hat{\sigma}_{i+1}^y)-\frac{h}{2}\sum_{i=1}^{N}\hat{\sigma}_i^z,
\end{equation}
where $J$ is the coupling constant, that from here on we assume to be equal to 1, $h$ the transverse field, $\hat\sigma^\alpha_i$ with $\alpha=x, y, z$ are the Pauli operators acting on the $i$-th spin and periodic boundary conditions are imposed $(\hat{\sigma}^\alpha_i=\hat{\sigma}^\alpha_{i+N})$. 
In the thermodynamic limit, the model in Eq.~\eqref{eq_XX} admits a first-order quantum phase transition that separates a gapped paramagnetic phase characterizing the region $|h|>2$, to a gapless critical phase for $|h| \leq 2$.
The second term is the translational symmetry-breaking term given by
\begin{equation}
\label{eq:pert}
\hat{H}_\text{def}=\frac{\varepsilon}{2}\sum_{j=-1}^1\hat{\sigma}^z_{k+jd},
\end{equation}
as depicted in Fig.~\ref{fig:chain}. 
In Eq.~\eqref{eq:pert}, $\varepsilon$ stands for the strength of the defects, and $d$ is the relative distance between them that we assume to be much smaller than the size of the system, i.e. $d \ll N$. 
The Hamiltonian in  Eq.~\eqref{eq_model} holds several symmetry properties. 
However, for our purposes, it is worth noting that $\hat{H}$ possesses a $U(1)$-symmetry, implying that the total magnetization along the $z$-axis is conserved, i.e. $[\hat{H},\sum_i\hat{\sigma}_i^z]=0$.

\begin{figure}
    \centering
    \includegraphics[width=\linewidth]{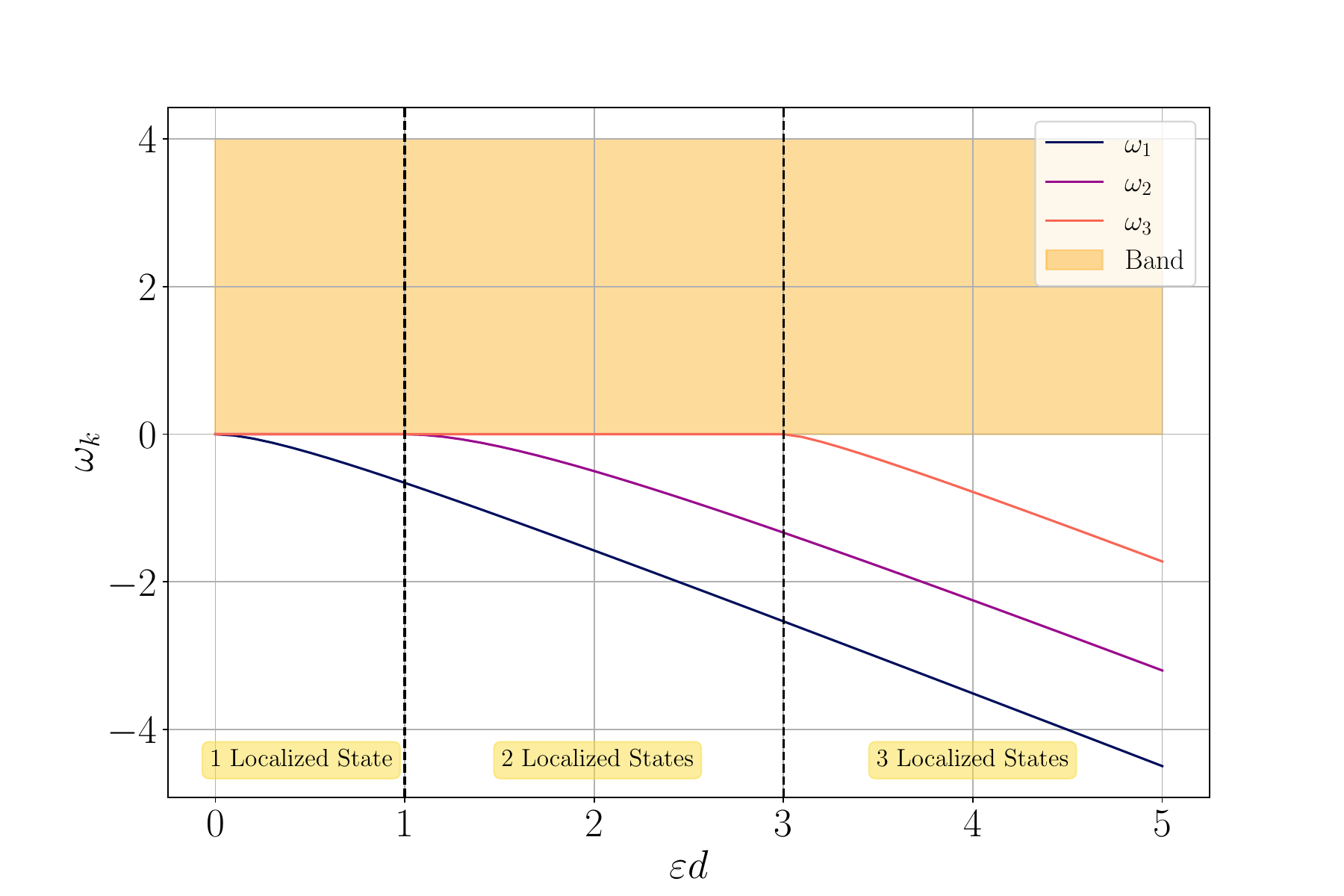}
    \caption{Single particle energy spectrum for $h=2$ illustrating the emergence of discrete energy levels below the continuous band as a function of $\varepsilon$ for $d=1$.}
    \label{fig:localised_h_2.0}
\end{figure}

Regardless of the presence of defects, the model in Eq.~\eqref{eq_model} can be mapped into a quadratic Hamiltonian of spinless fermions, with annihilation and creation operators given, respectively, by $\hat{c}_i, \hat{c}_i^{\dagger}$, via the Jordan-Wigner transformation~\cite{Lieb1961},
\begin{align}
\label{eq:hamFermions}
     H =~&J\sum_{i=1}^{N-1}(c_i^{\dagger}c_{i+1}+c_{i+1}^{\dagger}c_i)+\hat{P}_z\left(c_1^{\dagger}c_N+c_N^{\dagger}c_1\right) \nonumber \\
     &-h\sum_{i=1}^{N}c_i^{\dagger}c_i+\varepsilon\sum_{j=-1}^1c_{k+jd}^{\dagger}c_{k+jd}.
\end{align}
Here, $\hat{P}_z$ stands for the parity operator along $z$ with eigenvalues $\pm 1$.
Hence, depending on the total magnetization of the state, it induces antiperiodic or periodic boundary conditions~\cite{Damski2014, Franchini2016a}. 
In the following, we will consider only the positive parity subspace as the RDM of the three defects whose entanglement properties we investigate are equivalent to $N \gg 1$. 

\begin{figure}
    \centering
    \includegraphics[width=\linewidth]{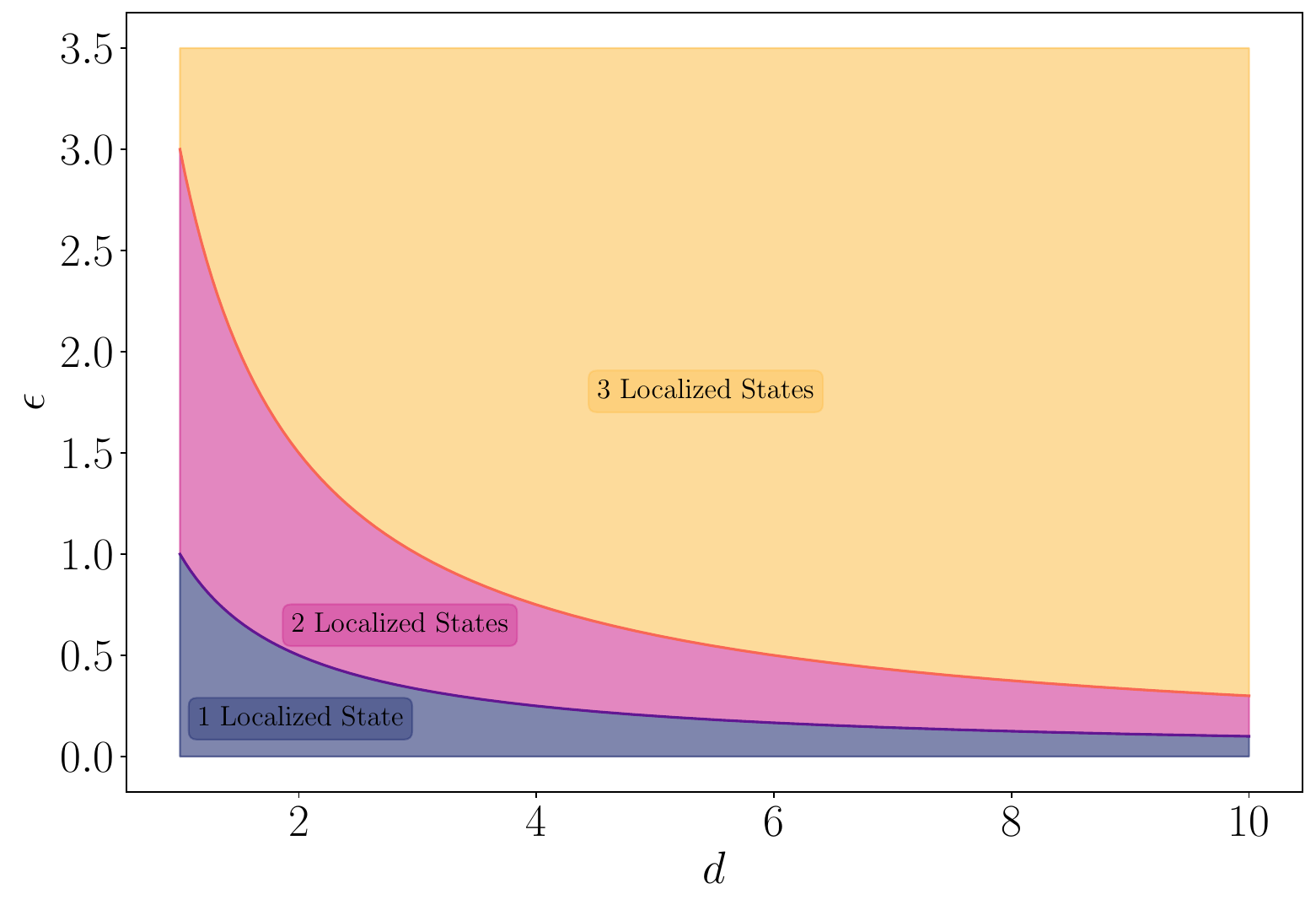}
    \caption{Showcasing the regions where one, two and three localized states appear in the $\varepsilon$-$d$ plane. The curve between the regions of one and two localized states is given by $\varepsilon = 1 / d$, and the curve between the regions of two and three localized states is given by $\varepsilon = 3 / d$.}
    \label{fig:defects}
\end{figure}

Because of its quadratic nature, the Hamiltonian in Eq.\eqref{eq:hamFermions} is readily diagonalized as
\begin{align}
    \label{eq:dia_Ham}
    \hat{H} = \sum_{k=1}^{N}\omega_k \hat{c}_k^{\dagger}\hat{c}_k,
\end{align}
where $\omega_k$ and $\hat{c}_k^{\dagger}=\sum_{n}\phi_{kn}\hat{c}_n$ are, respectively, the single-particle eigenenergies and eigenstates of the fermionic problem. 
In the limit $\varepsilon \rightarrow 0$ we recover $\omega_k=h-2\cos(2\pi k/N)$ and, in the thermodynamic limit, the single-particle spectrum makes up a continuous band with $\omega_k\in \left[h-2,h+2\right]$ with the eigenstates encoding plane waves~\cite{DePasquale2008}. 
In contrast, taking into account a strictly positive value of $\varepsilon$, up to three discrete energy levels emerge from below the band. 
Using a Green's function approach~\cite{Economou2006}, it is possible to analytically determine the boundaries $\varepsilon\left(d\right)$ between the regions with one, two, or three discrete eigenstates of the Hamiltonian. 
Those regions are:
\begin{subequations}
\label{eq:boundaries}
\begin{align}
    &0<\varepsilon<\frac{1}{d}, &&\text{one discrete energy level;} \\
    &\frac{1}{d}<\varepsilon<\frac{3}{d}, &&\text{two discrete energy levels;} \\
    &\frac{3}{d}<\varepsilon, &&\text{three discrete energy levels.}
\end{align}
\end{subequations}
Details of the derivation of these results can be found in Appendix~\ref{sec:GreenFTBH}. 
In Fig.~\ref{fig:localised_h_2.0} we show the energy spectrum of the fermionic Hamiltonian in Eq.~\eqref{eq:hamFermions}, with $h = 2$ and $J = 1$, as a function of the product $\varepsilon d$, where it is possible to have the emergence of discrete single-particle energies. 
Changing $h$, the single-particle energy spectrum is shifted vertically. 
Differently from the ones in the band, the eigenstates corresponding to the discrete eigenenergies are exponentially localized on the defects~\cite{Anderson1958}. 
In contrast, the continuous energy band gets distorted and their eigenstates acquire a contribution describing back-scattering at the impurity sites~\cite{Economou2006, Apollaro2008}.
From Eq.~\eqref{eq:boundaries} it is possible to identify three different regions in the $\varepsilon-d$ plane, each one of them characterized by a different number of localized states. These regions have been depicted in  Fig.~\ref{fig:defects}.

\section{Multipartite entanglement}
\label{sec:Multipartite entanglement}

As already mentioned, the main goal of this work is to study the multipartite entanglement in the ground state of the model in Eq.~\eqref{eq_model} between the spins on which magnetic defects are present. 
Toward this end, we first need to evaluate the RDM that is obtained by tracing out all the degrees of freedom associated with the spins on which no defects are present. 
The form of this matrix will take into account all the symmetries of the state from which it is obtained. 
In our case, it is well known that the ground state of the XX model with periodic boundary conditions, as in Eq.~\eqref{eq_XX}, has a vanishing momentum that implies that the RDM is real~\cite{Franchini2016a}. 
This result remains unaffected by the presence of defects. 
Moreover, since the distance between the first and second defect, and that between the second and the third are equal, there is a reflection symmetry with respect to the central defect. 
Furthermore, the Hamiltonian in Eq.~\eqref{eq_model} preserves the global magnetization and therefore all its eigenstates, including the ground-state, have a well-determined value of total magnetization. 
Taking into account all these properties the RDM in the computational basis becomes equal to
\begin{align}\label{eq_3den}
    \hat{\rho}=
    \begin{pmatrix}
       \rho_{00} & 0 & 0 & 0 & 0 & 0 & 0 & 0 \\
        0 & \rho_{11} & \rho_{12} & 0 & \rho_{14} & 0 & 0 & 0 \\
         0 & \rho_{12} & \rho_{22} & 0 & \rho_{12} & 0 & 0 & 0 \\
          0 & 0 & 0 & \rho_{33} & 0 & \rho_{35}  & \rho_{36}  & 0 \\
           0 & \rho_{14} & \rho_{12} & 0 & \rho_{11} & 0 & 0 & 0 \\
           0 & 0 & 0 & \rho_{35} & 0 & \rho_{55} & \rho_{35} & 0 \\
           0 & 0 & 0 & \rho_{36}  & 0 & \rho_{35} & \rho_{33} & 0 \\
           0 & 0 & 0 & 0 & 0 & 0 & 0 & \rho_{77} 
    \end{pmatrix}.
\end{align}

This matrix is, in general, a rank-8 density matrix that can be decomposed as follows:
\begin{align}\label{eq_opt_decom}
  \hat{\rho} =~&p_0 \op{000}{000}+\sum_{i=1}^3 p_i \op{gW_i}{gW_i}\nonumber\\
  &+\sum_{i=1}^3 \bar{p}_i \op{g\bar{W}_i}{g\bar{W}_i}+p_7 \op{111}{111},
\end{align}
where $\ket{gW_i}$ and $\ket{g\bar{W}_i}$ are, respectively, the generalized $W$- and the generalized spin-flipped $W$-state
\begin{align}
    \label{eq:gW}    
    &\ket{gW_i}=a^{(i)}_1\ket{001}+a^{(i)}_2\ket{010}+a^{(i)}_3\ket{100}\\    &\ket{g\bar{W}_i}=b^{(i)}_1\ket{011}+b^{(i)}_2\ket{101}+b^{(i)}_3\ket{110}.
\end{align}
This decomposition is particularly useful since it allows us to prove that the three-tangle~\cite{Coffman2000} of $\rho$ vanishes, i.e., $\tau_3\left( \rho\right)=0$. 
Such a result follows straightforwardly from the convex-roof extension of $\tau_3$ to mixed states~\cite{Horodecki2009} and from the well-known results that for a generalized $W$-state, $\tau_3$ is identically zero. 
This restricts the possibility of genuine multipartite entanglement to $W$-type~\cite{Acin2001}, which has $\tau_3=0$, but finite GME-concurrence $\mathcal{C}_\text{GME}$~\cite{Ma2011}.
The GME concurrence for pure states is defined as 
\begin{equation}
\label{eq:GME}
    \mathcal{C}_\text{GME}(\ket{\psi}) = \min_i \sqrt{2\left(1 - \Tr\left\{\rho_i^2\right\}\right)},
\end{equation}
where $\rho_i$ is the RDM of the $i$th bipartition. For mixed states, similar to the three-tangle, the GME concurrence can be calculated by exploiting the convex-roof extension
\begin{equation}
\mathcal{C}_\text{GME}\left(\rho\right) =  \inf \sum_i p_i \mathcal{C}_\text{GME}\left(\ket{\Psi_i} \right),
\end{equation}
such that $\rho=\sum_i p_i \op{\Psi_i}{\Psi_i}$ and the minimization procedure is carried out over all possible pure state decompositions of $\rho$. 

The computation of the GME concurrence for the case of more than one localized state (that is when the rank of the density matrix is equal or larger than two) is a highly non-trivial problem, which makes the analytical solution challenging to obtain. 
Consequently, we resort to numerical techniques to determine a lower bound for the GME concurrence exploiting two different methods; one presented by~\citet{Ma2011}; and the second by~\citet{Hong2012}. 
Both algorithms were embedded within a Monte Carlo algorithm using 100 runs with random initial states with the global minimum taken to be the one with the highest GME concurrence. 
The BFGS~\cite{Nocedal2006} optimizer was employed as the local optimizer, taken from the \texttt{SciPy} library~\cite{Virtanen2020}.

To complement the previous method, we also consider a separability criterion for the density matrix resorting to the algorithm presented by~\citet{Hofmann2014}, which iteratively attempts to decompose a given density matrix $\rho$ into a convex sum of biseparable components. 
At iteration $k+1$, the density matrix is expressed as the convex sum of $k$ terms $\op{\psi_k}$, where $\ket{\psi_k}$ is a biseparable state, and an additional semidefinite matrix $\rho_{k+1}$. 
If, for a certain $k$, the condition
\begin{equation}
\label{witness}
    \mathcal{W} = \Tr(\rho_{k+1}^2) - \frac{1}{7} < 0
\end{equation}
is satisfied, the density matrix is separable with respect to some bipartition (see \citet{Gurvits2002} for further details). 
The data is obtained by considering $1500$ iterations with $1000$ randomly generated biseparable states for each iteration.

Together with the multipartite entanglement, to have a complete characterization of $\rho$, we also have to evaluate the bipartite entanglement between each pair of spins. Such entanglement can easily be quantified by the concurrence~\cite{Wootters1998}. Due to the symmetry in the model, the concurrence between the first and the second must be equal to the one from the second and the third, i.e. $\mathcal{C}_{12} = \mathcal{C}_{23}$. Since the partial trace of $\rho$ with respect to any qubit is an X-state (that is, the RDM is zero excepts along the two diagonals), we can determine the concurrence via the simplified expression provided by \citet{Amico2004}. Accordingly, the concurrence between the first and second defect (and similarly between the second a third defect) is equal to
\begin{equation}\label{eq_conc12}
  \!\!\!\!\!\!  \mathcal{C}_{12}\! =\! 2\!\max\left\{\!0, |\rho_{12} \!+\! \rho_{35}|\! -\! \sqrt{(\rho_{00} \!+\! \rho_{44})(\rho_{33} + \rho_{77})}\!\right\}\!,
\end{equation}
while concurrence between the first and third defect is equal to
\begin{equation}\label{eq_conc13}
   \!\!\!\!\!\! \mathcal{C}_{13}\! =\! 2\!\max\left\{\!0, |\rho_{14} \!+\! \rho_{36}|\! -\! \sqrt{(\rho_{00}\! +\! \rho_{22})(\rho_{55}\! +\! \rho_{77})}\!\right\}\!.
\end{equation}

\section{Long-distance GME concurrence}
\label{Long-distance GME concurrence}

As discussed in Sec.~\ref{sec:model}, the XX model in Eq.~\eqref{eq_XX} exhibits two distinct thermodynamic phases, a gapless one for $|h| \leq 2$ and a gapped one for $|h|>2$. 
Therefore, it is natural to study the emergence of multipartite entanglement following the presence of defects in the two regions separately.

\subsection{The paramagnetic phase} \label{sec:h_2}

\begin{figure}
    \centering
    \includegraphics[width=\linewidth]{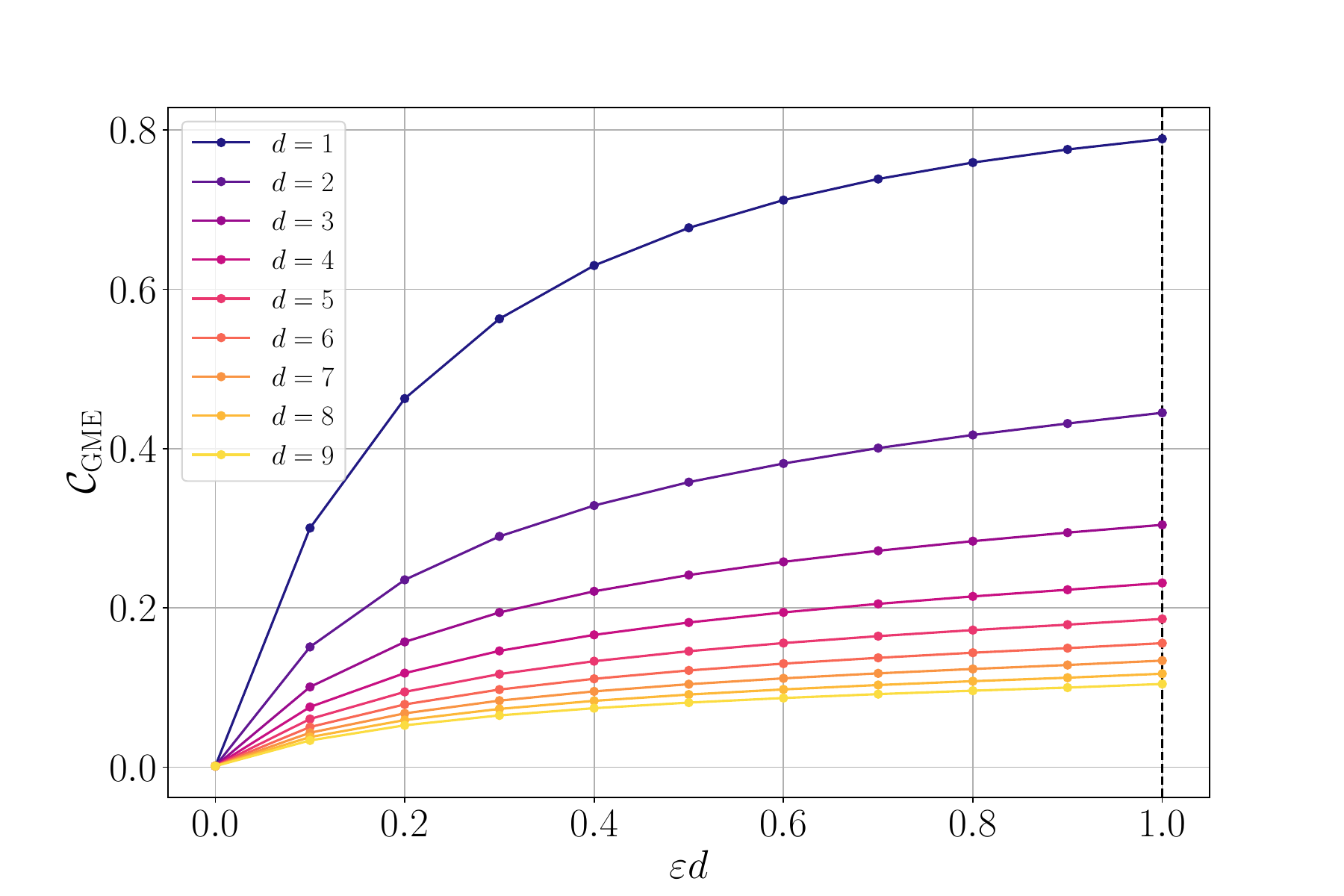}
    \caption{Analytically determined GME concurrence, as a function of $\varepsilon d$, in the presence of one localized state.}
    \label{fig:gme_h_2.0}
\end{figure}

For $h \geq 2$, discrete eigenstates emerge from below the continuous band for any $\varepsilon >0$ (see Fig.~\ref{fig:localised_h_2.0}). For $0 < \varepsilon < 1/d$, only one discrete energy level is present. 
As a consequence, the reduced density  matrix can be written as,
\begin{equation}
    \rho= p \op{gW} + (1 - p)\op{000},    
\end{equation}
which is of rank 2. 
This case is of particular interest since we can analytically determine the GME concurrence following the path depicted in Refs.~\cite{Lohmayer2006, Eltschka2008}. 
We start by analyzing the GME concurrence of the family of pure states
\begin{equation}
    \label{eq:pure_states}
    \ket{p, \varphi} = \sqrt{p}\ket{gW} + e^{i \varphi}\sqrt{1 - p}\ket{000}, 
\end{equation}
where $0 \leq p \leq 1$ and $0 \leq \varphi \leq 2\pi$, can be considered free parameters. 
Using Eq.~\eqref{eq:GME}, and recalling the expression of the generalized W-state in Eq.~\eqref{eq:gW} the GME concurrence of these states are
\begin{align}
    \label{eq:GME_pure_states}
    \mathcal{C}_\text{GME}(p, \varphi) = 2p &\min\left\{ |a_1|\sqrt{1 - |a_1|^2}, |a_2|\sqrt{1 - |a_2|^2}, \right. \nonumber \\
    &\hspace{0.835cm}\left. |a_3|\sqrt{1 - |a_3|^2} \right\}.
\end{align}
As a result, the zero polytope~\cite{Osterloh2008} of the states in Eq.~\eqref{eq:GME_pure_states} is a trivial one, where $p = 0$, assuming $|a_1|, |a_2|, |a_3| \neq 0, 1$. 
The next step is to construct the convex characteristic curve~\cite{Osterloh2008}. 
Since $0 \leq |a_1|, |a_2|, |a_3|  \leq 1$, then $\mathcal{C}_\text{GME}(p, \varphi)$ is monotonically increasing with $p$, and invariant with respect to $\varphi$. 
Moreover, $\mathcal{C}_\text{GME}(p, \varphi)$ is also linearly increasing with $p$, which implies that it is already convex. 
As a result, the GME concurrence is equal to the weight of the generalized $W$ state, similar to how the concurrence between a Bell state and an unentangled state is equal to the weight of the Bell state~\cite{Abouraddy2001}. 
Therefore
\begin{align}
    \mathcal{C}_\text{GME}(\rho) = 2p &\min\left\{ |a_1|\sqrt{1 - |a_1|^2}, |a_2|\sqrt{1 - |a_2|^2}, \right. \nonumber \\
    &\hspace{0.835cm}\left. |a_3|\sqrt{1 - |a_3|^2} \right\}.
\end{align}
In our case, due to the symmetry properties of our system, $a_3$ is equal to $a_1$ and $a_2 = \sqrt{1 - 2a_1^2}$, with $a_1 \in \left[0, 1/\sqrt{2}\right]$. 
Thus, the GME concurrence reduces to
\begin{align}\label{analytical_GME_one_defect}
    \mathcal{C}_\text{GME}(\rho) &= 2a_1 p \min\left\{\sqrt{1 - a_1^2}, \sqrt{2 - 4a_1^2}\right\} \nonumber \\
    &= 2\min\left\{\sqrt{2}|\rho_{12}|, \sqrt{|\rho_{14}|(1 - \rho_{00} - |\rho_{14}|)}\right\}. 
\end{align}
It has to be noted that, due to the range of values of $a_1$, $\rho$ always has a non-zero GME concurrence independently of the distance $d$, although it becomes vanishingly small as the distance increases. 
This dependence on the distance can be appreciated in Fig.~\ref{fig:gme_h_2.0} where we show the GME concurrence $\mathcal{C}_\text{GME}$ as a function of the rescaled defects' intensity $\varepsilon d$ in the region with a single localized energy eigenstate.

\begin{figure*}
    \centering
    \includegraphics[width=\linewidth]{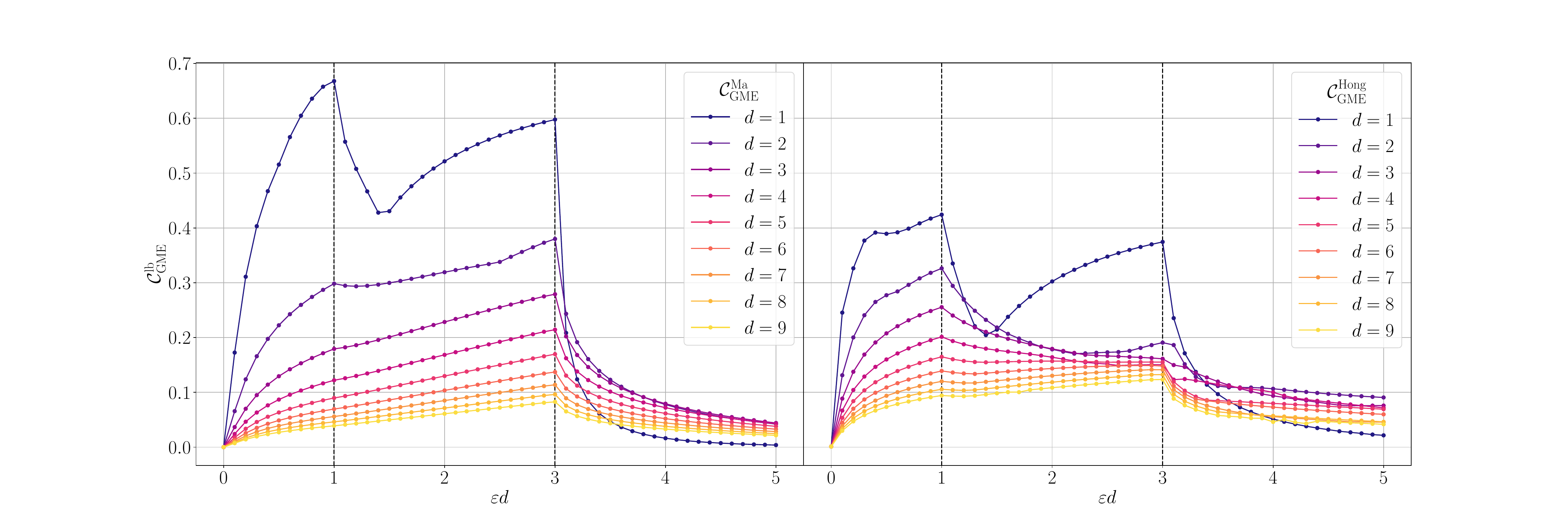}
    \caption{Lower bound on the GME concurrence as a function of $\varepsilon d$, for defects at a distance of up to 9, and $h = 2$. (left panel) Bound given by \citet{Ma2011}; (right panel) Bound given by \citet{Hong2012}. The vertical lines at $\varepsilon d=1$ and $\varepsilon d=3$, determine the boundaries between one and two, and between two and three localized eigenstates, respectively.}
    \label{fig:lb_gme_h_2.0}
\end{figure*}

For $\varepsilon > 1/d$, more localized states appear in the ground state of the system, and the rank of the RDM increases. 
This makes finding the convex-roof extension via analytical methods as done above for rank-2 density matrices highly non-trivial. 
Therefore, in the following, we resort to the numerical methods described in the previous section to determine the lower bound of the GME concurrence and corroborate our results via evaluating the separability criterion Eq.~\eqref{witness} of genuine multipartite entanglement.

\begin{figure}
    \centering
    \includegraphics[width=\linewidth]{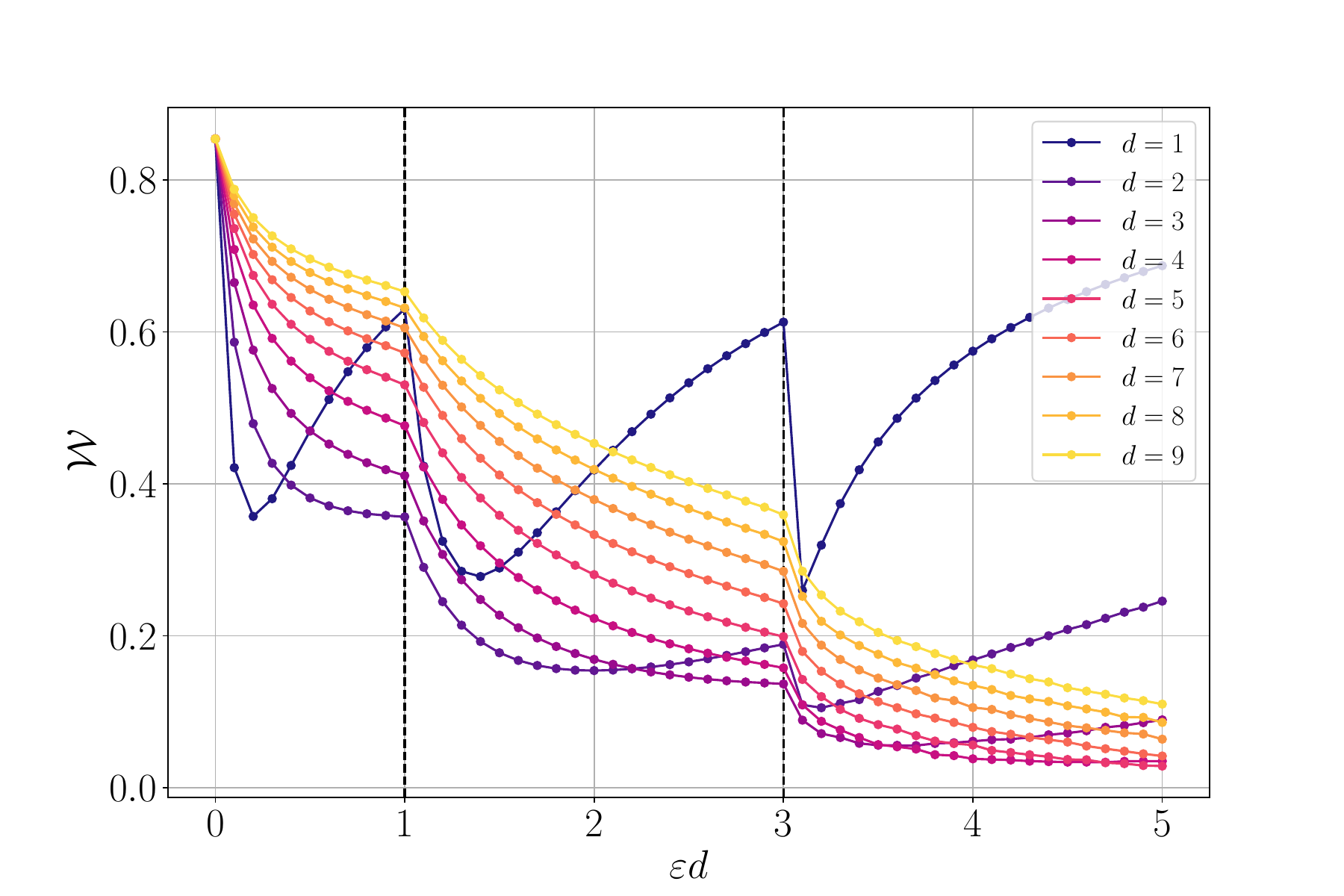}
    \caption{Witness $\mathcal{W}$ in Eq.(~\ref{witness}) as a function of $\epsilon d$  for $h=2$ and for different distances between the defects.}
    \label{fig:witness_h_2.0}
\end{figure}

Fig.~\ref{fig:lb_gme_h_2.0} shows the lower bound of the GME concurrence for $h = 2$ as a function of $\epsilon d$. 
For completeness, we compute this bound also for the previous case of only one localized eigenstate ($0 < \varepsilon d < 1$). 
We notice that, although the lower bounds of the GME concurrence decrease as the defects' separation $d$ increases, they remain finite indicating that long-distance GME concurrence is attainable for every value of $\varepsilon$ and $d$. 
Hence the numerically derived bounds for $\varepsilon d > 1$ confirm the presence of GME concurrence when multiple localized states contribute to the ground state. 
To strengthen these findings, we also consider the separability criterion Eq.~\eqref{witness}, which complements the GME results (Fig.~\ref{fig:witness_h_2.0}). 
We always found $\mathcal{W} \geq 0$, that is, the state is not separable for any bipartition. 
Finally, we also used the separability criterion based on the Hilbert-Schmidt distance developed in Ref.~\cite{Pandya2020} and strong numerical evidence suggests a non-zero distance to the closest bi-separable state. 
These results emphasize the ability of the defects to sustain long-distance multipartite entanglement that would otherwise be absent in uniform systems.

\begin{figure*}
    \centering
    \includegraphics[width=\linewidth]{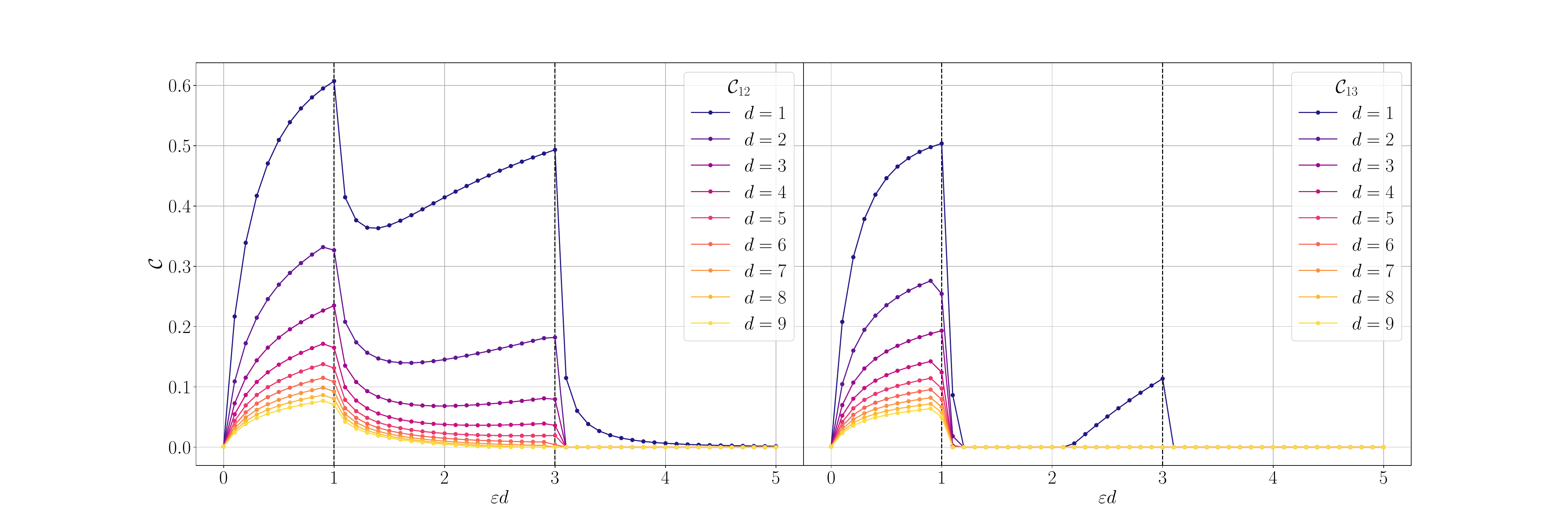}
    \caption{Two-qubit concurrence as a function of $\varepsilon d$, for defects at a distance of up to 9, and $h = 2$. (left panel) Concurrence between defects one and two (equivalent to the concurrence between defects two and three). (right panel) Concurrence between defects one and three. The vertical lines at $\varepsilon d=1$ and at $\varepsilon d=3$, determine the boundaries between one and two, and between two and three localized eigenstates, respectively.}
    \label{fig:conc_h_2.0}
\end{figure*}

In Fig.~\ref{fig:conc_h_2.0} we show the concurrence between the defects as evaluated via Eq.{~(\ref{eq_conc12})} and (\ref{eq_conc13}). 
While the concurrence between adjacent defects $\mathcal{C}_{12}$ and $\mathcal{C}_{23}$ is nonzero for smaller $\varepsilon d$, it vanishes beyond $\varepsilon d > 3$ and $d>1$. 
Similarly, the concurrence between the outer defects $\mathcal{C}_{13}$ is negligible for most of the parameter space. 
As a consequence, density matrices of the form of Eq.{~(\ref{eq_3den})} can sustain states with non-zero GME concurrence and vanishing two-qubit concurrence, similarly to $X$-type states~\cite{Rafsanjani2012}.

\subsection{The critical phase}

\begin{figure*}
    \centering
    \includegraphics[width=\linewidth]{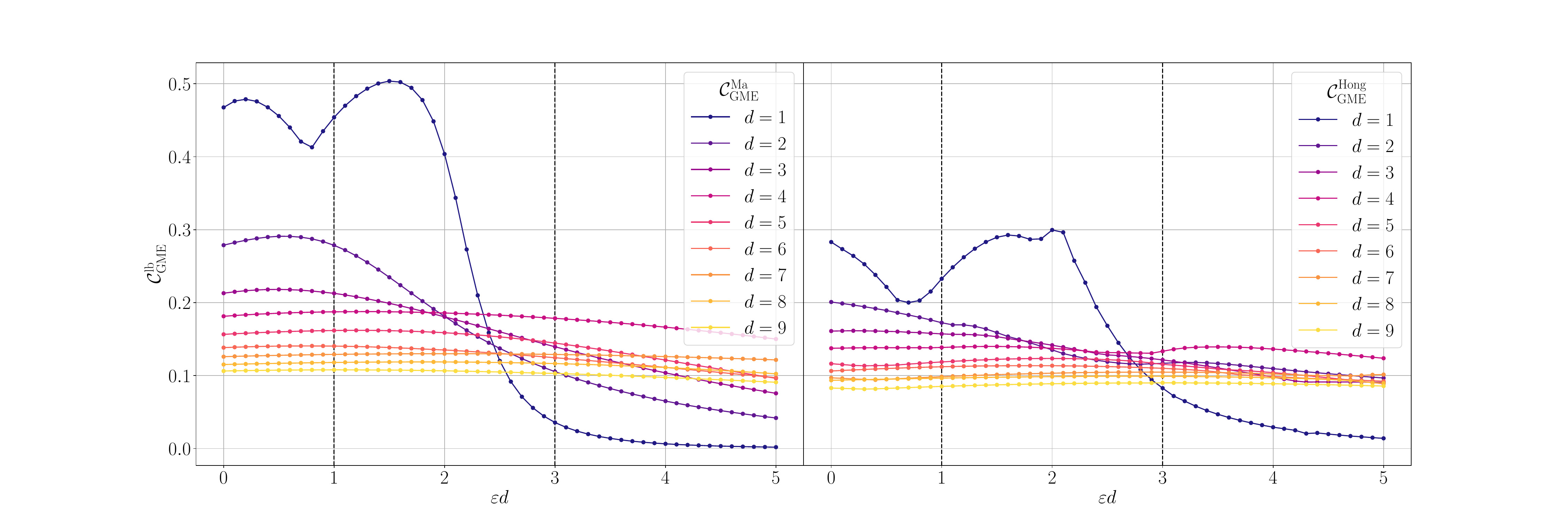}
    \caption{Lower bound on the GME concurrence as a function of $\varepsilon d$, for defects at a distance of up to 9, and $h = 1$. (left panel) Bound given by \citet{Ma2011}; (right panel) Bound given by \citet{Hong2012}. 
The vertical lines at $\varepsilon d=1$ and at $\varepsilon d=3$, determine the boundaries between one and two, and between two and three localized eigenstates, respectively.}
    \label{fig:lb_gme_h_1.0}
\end{figure*}

\begin{figure}
    \centering
    \includegraphics[width=\linewidth]{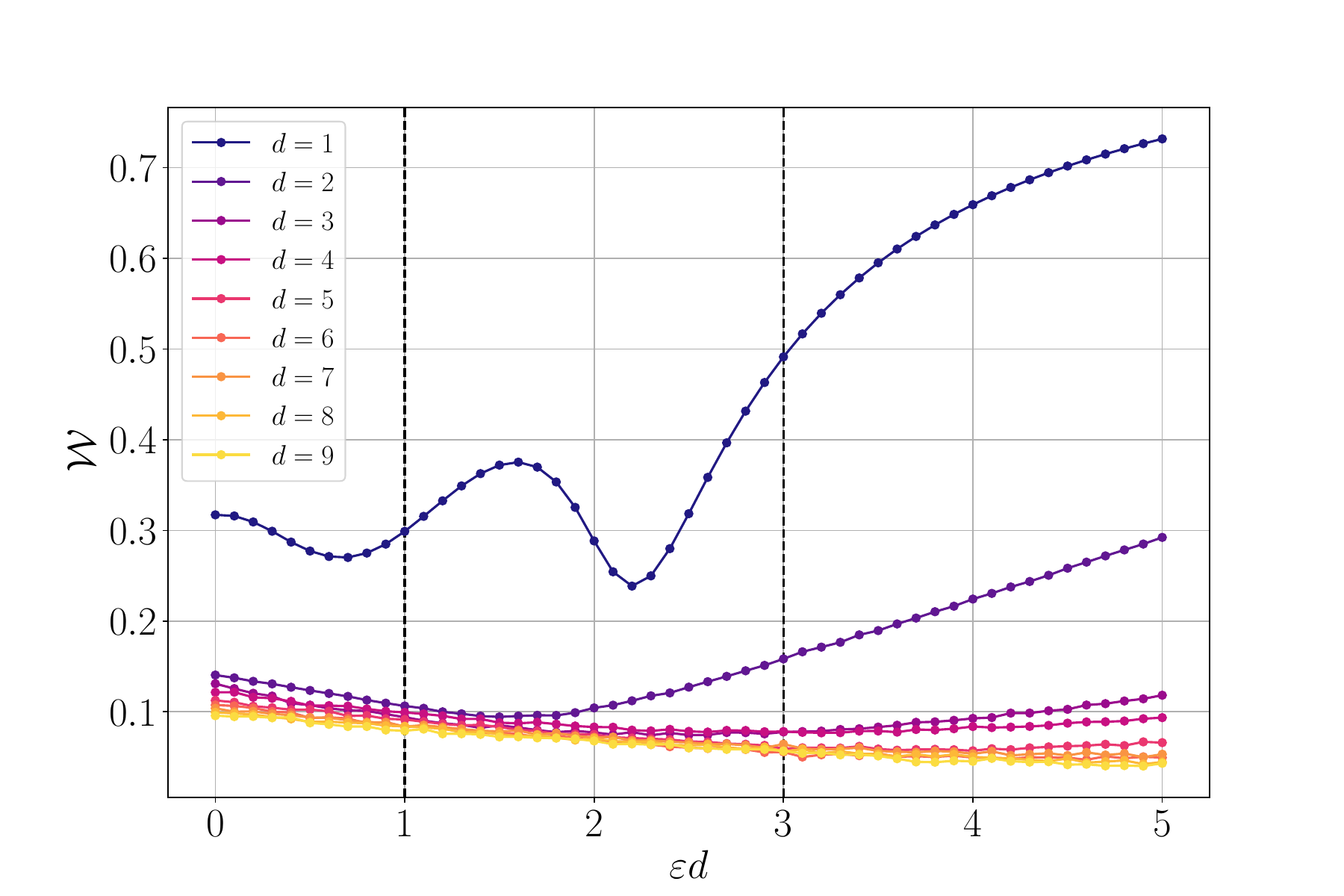}
    \caption{Witness $\mathcal{W}$ in Eq.(~\ref{witness}) as a function of $\epsilon d$  for $h=1$ and for different distances between the defects.}
    \label{fig:witness_h_1.0}
\end{figure}

For any $|h| \leq 2 $, regardless of the value of $\varepsilon \geq 0$, the RDM is always of rank 8, preventing an analytical determination of the GME concurrence. 
However, similar to Sec.~\ref{sec:h_2} above, we report the two lower bounds presented by \citet{Ma2011} and \citet{Hong2012}, the respective concurrences $\mathcal{C}_{12}$ and $\mathcal{C}_{23}$, as well as the separability criterion in Eq.~\eqref{witness}. Fig.~\ref{fig:lb_gme_h_1.0} shows the lower bounds for the GME concurrence and Fig.~\ref{fig:witness_h_1.0} the separability witness for $h = 1$. Similarly to the case of $h=2$, non-zero long-distance entanglement is present for any value of $\varepsilon d$ also in the absence of any pairwise concurrence between the qubits, which vanishes for $d > 1$ and $\varepsilon > 0$.

\section{Conclusion} \label{sec:conclusions}

In this work, we studied the interplay between magnetic defects and multipartite entanglement in an XX spin-$1/2$ chain subjected to a transverse field. 
Using analytical techniques complemented by numerical methods, we characterized the emergence of localized bound states at the defect sites.

Our findings demonstrate the existence of regions in parameter space, defined by the defect intensity and separation, where GME arises. 
For a single localized state, the GME concurrence was derived analytically. 
When multiple localized states are present, numerical approaches were employed to establish lower bounds on GME concurrence and validate multipartite entanglement through separability witnesses.

Interestingly, while pairwise entanglement between defects vanishes in certain regions, the GME concurrence persists, underscoring the robustness of multipartite entanglement at larger defect separations. 
This allows for the generation of long-distance multipartite entanglement in many-body system with nearest-neighbor interaction by the application of local magnetic defects at desired locations.
This approach to the generation of multipartite entanglement appears, simultaneously, easier and more flexible than the one based on models with Cluster interactions~\cite{Giampaolo_2014,Giampaolo_2015} hence disclosing the way of possible technological applications in quantum communication and quantum technologies.
Apart from such applications, our analytic results and numerical lower bounds shed light on the properties of multipartite entanglement in density matrices without coherence between different magnetization sectors, where only $W$-type entanglement can be present.

Although we restricted our analysis to three defects, the proposed method for long-distance multipartite entanglement generation via local magnetic fields can be straightforwardly extended to a higher number of defects to investigate higher SLOCC entanglement classes~\cite{Verstraete2002} or to the XY model, which sustains GHZ-type entanglement~\cite{Giampaolo2013}.

\acknowledgments

MC acknowledges funding by the Tertiary Education Scholarships Scheme MFED 758/2021/34. MC and TJGA acknowledge Xjenza Malta for their support via the project GROUNDS IPAS-2023-059. JO acknowledges the support by the PNRR MUR Project No. PE0000023-NQSTI, and thanks Viktor Eisler for useful discussions. RB acknowledges support from the Croatian Science Foundation (HrZZ) Project No. IP-2019-4-3321 and Austrian Science Fund (FWF) Grant-DOI:10.55776/P35434. MW acknowledges the support by NCN (Poland) Grant 2017/26/E/ST2/01008.

\appendix

\section{Green functions method} \label{sec:GreenFTBH}

We consider the $1D$ tight-binding Hamiltonian in the presence of three impurities, corresponding to the one in Eq.\eqref{eq:hamFermions}, decomposed as
\begin{equation}
    \label{eq:H3}
    H=H_0+H_1,
\end{equation}
where $H_0$ is the tight-binding Hamiltonian without impurities, characterized by nearest-neighbor interactions
\begin{equation}
    \label{eq:TBH0}
    H_0=\sum_i h\op{i}+J\sum_i(\op{i}{i+1}+\text{h.c.}),
\end{equation}
and $H_1=H_l+H_m+H_n$ denotes the perturbation, that is assumed to be diagonal,
\begin{equation}
    H_1=\varepsilon_l\op{l}+\varepsilon_m\op{m}+\varepsilon_n\op{n}.
\end{equation}
We introduce the unperturbed Green operator $G_0$  corresponding to $H_0$,
\begin{equation}
    \label{eq:G0def}
    G_0(z)=\frac{1}{z-H_0}=\sum_k\frac{\op{k}}{z-E_k}.
\end{equation}
In the limit $N\rightarrow \infty$, the set of states $\ket{k}$ form a continuous band with energies $E_k\in I_b=[2h-J, 2h+J]$. The matrix elements of $G_0$ between two localized states are given by
\begin{equation}
    G_0(r,s;z)=\frac{(-x+ \sqrt{x^2-1})^{|r-s|}}{J\sqrt{x^2-1}}, \quad z\notin I_b,
\end{equation}
and
\begin{equation}
    G_0(r,s;z)=\frac{(-x\pm \imath \sqrt{1-x^2})^{|r-s|}}{\pm \imath J\sqrt{x^2-1}}\quad z\in I_b,
\end{equation}
where $x=\frac{z-2h}{J}$. The full Green operator $G$, corresponding to the full Hamiltonian $H$, can be obtained as
\begin{equation}
    \label{eq:Gpert}
    G=G_0+G_0TG_0,
\end{equation}
where the $T$-matrix is determined from the knowledge of $G_0$ and $H_1$, as 
\begin{equation}
    \label{eq:Tmatrix}
    T=H_1+H_1G_0H_1+H_1G_0H_1G_0H_1+\cdots.
\end{equation}
The $T$-matrix can be obtained following the approach in~\cite{Economou2006}. Introducing the unit operator into Eq.\eqref{eq:Tmatrix}, we get
\begin{equation}
    T=H_1+H_1\sum_i\op{i}G_0\sum_{i'}\op{i'}H_1+\cdots.
\end{equation}
Since the defects are diagonal, we need to retain in the summation only the terms $(\op{l}+\op{m}+\op{n})$, that can be written as scalar product $\op{\alpha}$, with
\begin{equation}
    \ket{\alpha}\equiv[\ket{l}, \ket{m}, \ket{n}],~\bra{\alpha}\equiv 
    \begin{bmatrix} 
        \bra{l} \\
        \bra{m} \\
        \bra{n}
    \end{bmatrix}.
\end{equation}
With this notation, we have
\begin{equation}
    \label{eq:Tmat}
    \bra{\alpha}T\ket{\alpha}=\bra{\alpha}H_1\ket{\alpha}(\mathbb{I}_3-\bra{\alpha}G_0\ket{\alpha}\bra{\alpha}H_1\ket{\alpha})^{-1},
\end{equation}
where
\begin{align}    \bra{\alpha}H_1\ket{\alpha}&=\begin{pmatrix}
        \varepsilon_l & 0 & 0\\
        0 & \varepsilon_m & 0\\
        0 & 0 & \varepsilon_n
    \end{pmatrix}, \\
      \bra{\alpha}G_0\ket{\alpha}&=\begin{pmatrix}
        G_0(l,l) & G_0(l,m) & G_0(l,n)\\
        G_0(m,l) & G_0(m,m) & G_0(m,n)\\
        G_0(n,l) & G_0(n,m) & G_0(n,n)
    \end{pmatrix}.
    \end{align}
The first step is to invert the matrix
\begin{align}
&\mathbb{I}_3-\bra{\alpha}G_0\ket{\alpha}\bra{\alpha}H_1\ket{\alpha}= \nonumber \\ &\begin{pmatrix}
        (1-\varepsilon_lG_0(l,l)) & -\varepsilon_mG_0(l,m) & -\varepsilon_n G_0(l,n)\\
        -\varepsilon_lG_0(m,l) & (1-\varepsilon_mG_0(m,m)) & -\varepsilon_n G_0(m,n)\\
        -\varepsilon_lG_0(n,l) & -\varepsilon_mG_0(n,m) & (1-\varepsilon_nG_0(n,n))
    \end{pmatrix}.
\end{align}
The determinant gives
\begin{align}
&   \det[\mathbb{I}_3-\bra{\alpha}G_0\ket{\alpha}\bra{\alpha}H_1\ket{\alpha}]= \nonumber\\
&   f_{lmn}^{-1}(1-\varepsilon_lG_0(l,l))(1-\varepsilon_mG_0(m,m))(1-\varepsilon_nG_0(n,n)),
\end{align}
where we define $f_{lmn}$ as
\begin{align}
\label{eq:flmn}
  &  f_{lmn}=[1-t_lt_mG_0(l,m)G_0(m,l)-t_mt_nG_0(m,n)G_0(n,m) \nonumber \\
    &-t_nt_lG_0(n,l)G_0(l,n)
    -t_lt_mt_n(G_0(l,m)G_0(m,n)G_0(n,l) \nonumber \\
    &+G_0(m,l)G_0(l,n)G_0(n,m))]^{-1},
\end{align}
and
\begin{equation}
\label{eq:tj}
    t_j=\frac{\varepsilon_j}{1-\varepsilon_{j}G_0(j,j)},~j=l,m,n.
\end{equation}
The poles of the Green functions for the three-defects problem can be determined from the zeroes of Eq.~\eqref{eq:flmn}. In the case $\varepsilon_l = \varepsilon_m = \varepsilon_n$ and $\left|l-m\right|=\left|n-m\right|=d$, that is, the intensity of the magnetic field on the three defects is the same and the defects are symmetrically placed around the central defect, as in Fig.~\ref{fig:localised_h_2.0} of the main paper, we find, using the Mathematica software package and confirmed through numerical simulations, that for $0<\varepsilon<1/d$ only one solution is found, for $1<\varepsilon<3/d$ two solutions are found, and for $3/d<\varepsilon$, three solutions are found, as reported in Eqs.~\eqref{eq:boundaries} in the main paper.

The emergence of three localized states can also be obtained by solving the 1D Schrodinger equation with three delta potentials:
\begin{equation}
     -\frac{d^2 \psi(x)}{dx^2} + V(x)\psi(x) = E\psi(x),
\end{equation}
where the potential is
\begin{equation}
    V(x) = -\varepsilon\left[\delta(x + d) + \delta(x) + \delta(x - d)\right].
\end{equation}
Straightforward calculations give that the solutions for bound states of this problem are given as:
\begin{subequations}
\begin{align}
    \text{1 even solution: } &0 < d < \frac{\hbar^2}{2m\varepsilon}, \\
    \text{1 odd and 1 even solution: } &\frac{\hbar^2}{2m\varepsilon} < d < \frac{3\hbar^2}{2m\varepsilon}, \\
    \text{1 odd and 2 even solutions: } &\frac{3\hbar^2}{2m\varepsilon} < d,
\end{align}
\end{subequations}
and by setting $\hbar^2/2m = 1$, we obtain Eqs.~\eqref{eq:boundaries}.

\bibliographystyle{apsrev4-2}
\bibliography{ref}

\end{document}